\documentclass[aps,twocolumn,showpacs]{revtex4}

\usepackage{amsmath}
\usepackage{graphicx}

\begin{document}

\title{Nonlinear graphene plasmonics: amplitude equation}
\author{A.V. Gorbach}
\affiliation{Centre for Photonics and Photonic Materials, Department of Physics,
University of Bath, Bath BA2 7AY, United Kingdom}

\begin{abstract}
Using perturbation expansion of Maxwell equations, the amplitude equation is derived for nonlinear TM and TE surface plasmon waves supported by graphene. The equation describes interplay between in-plane beam diffraction and nonlinerity due to light intensity induced corrections to graphene conductivity and susceptibility of dielectrics.
For strongly localized TM plasmons, graphene is found to bring the superior contribution to the overall nonlinearity. In contrast, nonlinear response of the substrate and cladding dielectrics can become dominant for weakly localized TE plasmons.
\end{abstract}

\pacs{42.65.Wi; 78.67.Wj; 73.25.+i; 78.68.+m}

\maketitle

\section{Introduction}

Applications of graphene in photonics and optoelectronics are being actively discussed in recent years
\cite{Bonaccorso2010, Bao2012}. In particular, graphene plasmonics is considered as a promissing alternative 
to conventional plasmonics with noble metals 
\cite{Koppens}. Recently, hybrid metal-graphene plasmonic structures have been proposed
as the propitious platform for novel optical devices \cite{Grigorenko2012}.
 
Graphene supports two types of surface plasmons: transverse magnetic (TM) and transverse electric (TE) modes \cite{Mikhailov2007a,Jablan2009}. 
TM graphene plasmon is in many ways analogous to the surface plasmon excited
at a metal/dielectric interface \cite{book_Maier}, 
although specific features of collective electron excitation in the purely 2D graphene lead to
qualitative differences in the spectra of plasmons in these two systems \cite{Jablan2009}.
Compared to its metal analogue, TM plasmon supported by graphene offers susbstantial enhancement of the field localization, accompained by the considerable decrease of the propagation loss -- all being crucial for
potential applications of surface plasmons in miniature photonic components.
The existence of TE plasmon is directly related to the linear (Dirac) spectrum of electrons in graphene \cite{Mikhailov2007a}, there is no analogue 
of such surface wave in conventional plasmonics. TE plasmon is only weakly localized at the surface, 
however it is characterized by considerably low propagation losses even at room temperatures.
Spectral characteristics of TE and TM graphene plasmons are defined by the charge density, which can be controlled chemically \cite{Mak2008} or electrically \cite{Chen2011}. This tunability represents another important advantage of graphene plasmons over metal plasmons.

Optical properties of doped graphene are encapsulated in the induced surface current $\mathcal{K}$.
So far, graphene plasmons have been studied under the assumption of a linear relation between the current $\mathcal{K}$ and the field amplitude $\mathcal{E}$: $\mathcal{K}=\sigma \mathcal{E}$ \footnote{In contrast to standard conductivity for metals, here $\sigma$ is related to the full current $\mathcal{K}$ in graphene layer, and not the current density $\mathcal{J}$. It is therefore measured  is $S=A/V$.}. This is true only at low light intensities, while 
generally the dependence $\mathcal{K}(\mathcal{E})$ is predicted to be highly nonlinear \cite{Mikhailov2007, Mikhailov2008}. The particularly strong nonlinear repsonse of graphene has been confirmed in several experiments, including direct measurements with optical Kerr gate \cite{Chu2011} and z-scan \cite{Zhang2012} techniques, as well as
observation of four-wave mixing with graphene flakes \cite{Hendry2010} and a range of nonlinear effects in a graphene-coated photonic crystal 
nano-cavity \cite{Gu2012}. Altogether these findings put forward the great potential of graphene for building functional nano-photonic devices.

\begin{figure}
\includegraphics[width=0.45\textwidth]{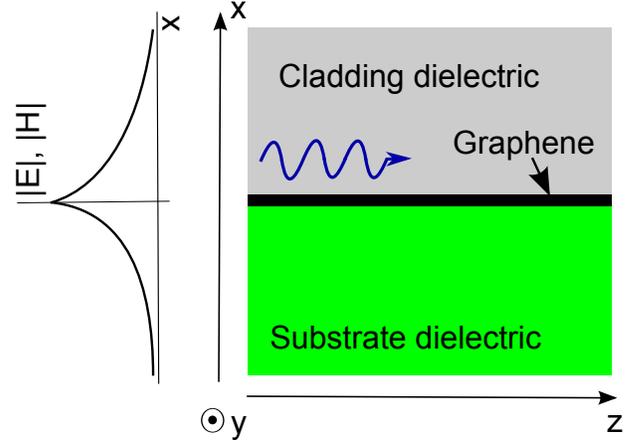}
\caption{(Color online) Schematic illustration of surface plasmon propagating along graphene sheet. Fields are exponentionally localized in $x$ (across the interface), as shown in the left panel.}
\label{fig:geom}
\end{figure}

In this work we consider nonlinear surface waves supported by graphene in the simple planar geometry shown in Fig.~\ref{fig:geom}. 
Allowing for light intensity corrections to the surface current and to the
susceptibility of dielectrics surrounding graphene, as well as introducing diffraction due to a finite beam width in the unbound ($y$) direction, we develop asymptotic expansion of Maxwell equations and boundary conditions to obtain an amplitude equation for quasi-TM and quasi-TE surface waves. 
The asymptotic expansion procedure is similar to that recently developed for semiconductor and metal nano-waveguides \cite{SGM2011,Marini2011}.
Further, we analyze the relative contribution from dielectrics and graphene to the overall effective nonlinearity of the system for the two types of plasmons, and 
the impact of geometry on the nonlinearity enhancement. 

\section{Setup and asymptotic expansion of Maxwell equations}

We consider the planar geometry, in which single layer graphene is sandwiched in-between two dielectrics. We choose $x$ axis to be perpendicular to the interfaces, $z$ is the direction of propagation, and $y$ is the unbound direction in which light can diffract, see Fig.~\ref{fig:geom}. For monochromatic fields, $\mathcal{\vec{E}}=\frac12 \vec{E} e^{-i\omega t} + c.c.$, $\mathcal{\vec{H}}=\frac12 \vec{H} e^{-i\omega t} + c.c.$, in each dielectric domain we solve stationary Maxwell equations:
\begin{equation}
\vec{\nabla}\times\vec{\nabla}\times \vec{E}= \frac{\vec{D}}{\epsilon_0}\;.
\label{eq:Maxwell}
\end{equation}
Here spatial coordinates are normalized to the inverse wave number $k=2\pi/\lambda=\omega/c$.
For homogeneous isotropic dielectrics, dispacement vector takes the form:
\begin{eqnarray}
\vec{D}&=&\epsilon_0\left[\epsilon \vec{E}+\vec{N}\right]\;,\\
\vec{N}&=&\frac12\chi_3\left(|\vec{E}|^2 \vec{E}+\frac12 \vec{E}^2\vec{E}^*\right)\;.
\label{eq:Kerr}
\end{eqnarray}

Adapting complex amplitude notation to the surface current $\mathcal{\vec{K}}=\frac12 \vec{K} e^{-i\omega t} + c.c.$, $\vec{K}=[0,K_y,K_z]^T$, 
the boundary conditions can be written as:
\begin{eqnarray}
\label{eq:BC_Ey_and_Ez}
\Delta[E_y]=0\;, \qquad \Delta[E_z]=0\;,\\
\label{eq:BC_Hy}
-\Delta[H_{y}]=ic\epsilon_0\Delta[\partial_z E_x-\partial_x E_z]=K_z\;,\\
\label{eq:BC_Hz}
-\Delta[H_{z}]=ic\epsilon_0\Delta[\partial_x E_y-\partial_y E_x]=-K_y\;,
\end{eqnarray}
where operators $\Delta$ and $\Theta$ are defined as:
\begin{eqnarray}
\Delta[f(x)]=\lim_{\delta\to 0}\left(f(x-\delta)-f(x+\delta)\right)\;,\\
\Theta[f(x)]=\frac12\lim_{\delta\to 0}\left(f(x-\delta)+f(x+\delta)\right)\;.\;
\end{eqnarray}

Taking into accound first order nonlinear corrections to the relationship betwen surface current and electric field amplitude $\mathcal{\vec{K}}(\mathcal{\vec{E}})$ \cite{Mikhailov2008} and neglecting the effect of higher harmonics generation, we obtain:
\begin{equation}
K_{y,z}=\Theta\left[\sigma_1 E_{y,z}+\frac{\sigma_3}{2} \left( |\vec{E}|^2 E_{y,z}+\frac12\vec{E}^2 E^*_{y,z}\right)\right]\;,
\label{eq:surf_current}
\end{equation}

If the nonlinear response is neglected altogether ($\sigma_3=0$, $\chi_3=0$), and no losses due to electron-phonon scattering or defects are considered at zero temperature ($Re(\sigma_1)=0$), the above system admits solutions in the form of surface plasmons propagating in $z$ direction: $E,H\sim e^{i\beta z}$. Field amplitudes in such solutions are exponentially localized at the interface $x=0$ and constant along the unbound direction $y$. For the case of positive/negative imaginary part of conductivity $\sigma_1$ only TM/TE surface plasmon exists \cite{Mikhailov2007a}.

Below we consider a weakly dissipative case, so that $\sigma_1=\sigma_1^{(R)}+i\sigma_1^{(I)}$ and $\sigma_1^{(R)}/\sigma_1^{(I)}\sim s \ll 1$, where $s$ is a dummy small parameter. This assumption is valid for a highly doped graphene, $|\mu|\gg kT$ and $\hbar\omega<2|\mu|$, $\mu$ is the chemical potential  \cite{Mikhailov2007a}. Furthermore, we 
assume that nonlinear corrections to the dielectric susceptibility $\sim\chi_3|\vec{E}|^2$ and graphene conductivity $\sim\sigma_3|\vec{E}|^2$ are of the same order of smallness $O(s)$. We let the mode amplitude $\psi$ to vary slowly with propagation distane, $\partial_z \psi \ll \beta\psi$, and consider weak diffraction, $\partial_y\psi\ne 0$. Using asymptotic expansion of Maxwell Eqs.~(\ref{eq:Maxwell}) and boundary conditions, Eqs.~(\ref{eq:BC_Ey_and_Ez})-(\ref{eq:BC_Hz}), below we derive propagation equation for the mode amplitude $\psi$. 

Note, with the account of diffraction, the separation into TM and TE modes can no longer be performed, instead one deals with quasi-TM and quasi-TE modes.

\section{Quasi-TM surface plasmon}

We seek a guided mode solution in the form:
\begin{eqnarray}
E_x&=&\left[A_x(\psi,x) +B_x(\psi,x)+O(s^{5/2})\right]e^{i\beta z}\;,\\
E_y&=&\left[C(\psi,x)+O(s^2)\right]e^{i\beta z}\;,\\
E_z&=&\left[A_z(\psi,x) +B_z(\psi,x)+O(s^{5/2})\right]e^{i\beta z}\;,
\end{eqnarray}
where $\psi=\psi(z,y)$ is a slowly varying function: $\partial_z\psi\sim s$, $\partial_y\psi\sim s^{1/2}$, $A_{x,z}\sim s^{1/2}$, $C\sim s$, $B \sim s^{3/2}$. The chosen orders of smallness are justified below by solving consistently boundary value problems, that emerge in different orders of $s$.
Following substitution into Maxwell equations, in the order $O(s^{1/2})$ we obtain the following boundary value problem:
\begin{eqnarray}
\hat{L}_{TM}\vec{A}&=&0\;, \\ 
\label{eq:BC_order_12}
\Delta[A_z]=0\;, && \Delta[i\beta A_x - \partial_x A_z]=\alpha_1^{(I)} \Theta[A_z]\;,
\end{eqnarray}
where $\vec{A}=[A_x,A_z]^T$, $\alpha_1=\sigma_1/(c\epsilon_0)$, and operator $\hat{L}_{TM}$ is defined as:
\begin{equation}
\label{eq:L_TM}
\hat{L}_{TM}=\left[
\begin{array}{cc}
\beta^2-\epsilon & i\beta\partial_x \\
i\beta\partial_x & -\partial_{xx}^2-\epsilon
\end{array}
\right]\;.
\end{equation}

We choose the solution in the form $\vec{A}=I^{1/2}\psi(z,y)\vec{e}$, where $\vec{e}=[e_x,e_z]^T$ is the linear surface plasmon mode:
\begin{eqnarray}
x<0&:& e_z=e^{q_sx}\;,\qquad e_x=\frac{-i\beta}{q_s}e^{q_s x}\\
x>0&:& e_z=e^{-q_cx}\;,\qquad e_x=\frac{i\beta}{q_c}e^{-q_c x}\\
\label{eq:q_sc}
&&q_{s,c}=\sqrt{\beta^2-\epsilon_{s,c}}\;,
\end{eqnarray}
$\epsilon_{s}$ and $\epsilon_{c}$ correspond to dielectric layers at $x<0$ (substrate) and $x>0$ (cladding), respectively. Propagation constant $\beta$ is defined through
the dispersion relation \cite{Jablan2009}:
\begin{equation}
\frac{\epsilon_s}{\sqrt{\beta^2-\epsilon_s}}+\frac{\epsilon_c}{\sqrt{\beta^2-\epsilon_c}}=\alpha_1^{(I)}
\label{eq:disp_TM}
\end{equation}

The normalization factor $I$ is chosen in a way that
$|\psi|^2$ is the power density (measured in watts per meter) carried
in the $z$ direction \cite{SGM2011}:
\begin{eqnarray}
I&=&\frac{2\beta k}{\epsilon_0 c Q}\;,\\ 
Q&=&\int_{-\infty}^{+\infty} \epsilon |e_x|^2 dx=\frac{\beta^2}{2}\left(\frac{\epsilon_s}{q_s^3}+\frac{\epsilon_c}{q_c^3}\right)
\end{eqnarray}

Collecting terms of the order $O(s)$ we obtain:
\begin{eqnarray}
(\beta^2-\epsilon)C-\partial_{xx}^2 C = -I^{1/2}\partial_y\psi\left(i\beta e_z+\partial_x e_x \right)\;,\\
\label{eq:BC_order_s}
\Delta[C]=0\;, \qquad \Delta[\partial_x C - I^{1/2}\partial_y\psi e_x]=-\alpha_1^{(I)} \Theta[C]\;.
\end{eqnarray}
From $\vec\nabla\cdot \vec{E}=0$ in the order $O(s^{1/2})$ it follows that $i\beta e_z+\partial_x e_x=0$, and therefore $C$ solves the homogeneous equation. It is non-zero due to simultaneous diffraction ($\partial_y\psi\ne 0$) and discontinuity of $e_x$ component at the interface, see Eq.~(\ref{eq:BC_order_s}). Substituting $C=I^{1/2}\partial_y\psi e_y$, it is easy to see that $e_y$ satisfies the same homogeneous equation as $e_z$. Comparing boundary conditions for $e_y$ and $e_z$, we obtain $e_y=(-i/\beta) e_z$.

In the order $O(s^{3/2})$ we obtain the following boundary value problem:
\begin{eqnarray}
\label{eq:order_s32}
&\hat{L}_{TM}\vec{B}=I^{1/2}\vec{J}\;,&\\
&\Delta[B_z]=0\;,&\\
\nonumber
&\Delta[i\beta B_x+\partial_z \psi I^{1/2}e_x-\partial_x B_z]=-i\alpha_1^{(R)}I^{1/2}\psi\Theta[e_z]&\\
\label{eq:BC_order_s32}
&+\alpha_1^{(I)}\Theta[B_z]-\frac{i}{2} \alpha_3  I^{3/2} |\psi|^2\psi \Theta\left[|\vec{e}|^2 e_z + \frac12\vec{e}^2 e_z^*\right]\;,&
\end{eqnarray}
where $\alpha_3=\sigma_3/(c\epsilon_0)$ and
\begin{eqnarray}
\nonumber
J_x&=&\partial_z\psi(2i\beta e_x -\partial_x e_z)+\partial_{yy}^2\psi (e_x-\partial_x e_y)\\
&&+|\psi|^2\psi \; n_x\;,\\
\nonumber
J_z&=&-\partial_z \psi \partial_x e_x +\partial_{yy}^2\psi (e_z-i\beta e_y)\\
&&+|\psi|^2\psi \; n_z\;,
\end{eqnarray}
$n_{x,z}=IN_{x,z}(e_x,0,e_z)$.

Next, we project Eq.~(\ref{eq:order_s32}) onto the linear mode $\vec{e}$:
\begin{equation}
\label{eq:project}
\int_{-\infty}^{+\infty}\left(\vec{e}^*\cdot \hat{L}_{TM}\vec{B}\right)dx=I^{1/2}\int_{-\infty}^{+\infty}\left(\vec{e}^*\cdot\vec{J}\right)dx\;,
\end{equation}
take $\int_{-\infty}^{+\infty}=\int_{-\infty}^{0}+\int_{0}^{+\infty}$ in the l.h.s., apply integration by parts and use boundary conditions in 
Eqs.~(\ref{eq:BC_order_12}) and (\ref{eq:BC_order_s32}) to obtain:
\begin{eqnarray}
\nonumber
&&\int_{-\infty}^{+\infty}\left(\vec{e}^*\cdot \hat{L}_{TM}\vec{B}\right)dx=\\
\nonumber
&&\qquad\Delta\left[i\beta(e_{z}^*B_{x}+e_{x}^*B_{z})-e_{z}^*\partial_xB_{z}+B_{z}\partial_x e_{z}^*\right]\\
\nonumber
&&=-\partial_z\psi I^{1/2}\Delta[e_xe_z^*]-i\alpha_1^{(R)}\psi I^{1/2}\Theta[|e_z|^2]\\
&&\qquad-\frac{i}{2}\alpha_3|\psi|^2\psi I^{3/2}\Theta\left[|\vec{e}|^2|e_z|^2+\frac12\vec{e}^2(e_z^*)^2\right]\;.
\end{eqnarray}

Finally, computing integrals in the r.h.s. of Eq.~(\ref{eq:project}), we obtain the amplitude equation:
\begin{equation}
\label{eq:NLS}
i\frac{\partial\psi}{\partial(z/k)}+\frac{1}{2\beta k}\frac{\partial^2\psi}{\partial(y/k)^2}+i\Lambda \psi+\Upsilon|\psi|^2\psi=0\;,
\end{equation}
where the nonlinear parameter $\Upsilon$ combines contributions of graphene and dielectrics:
\begin{eqnarray}
\label{eq:Upsilon}
\Upsilon&=&g(\gamma_G+\gamma_D)\;,\\
\label{eq:gamG}
\gamma_G&=&\frac{i\alpha_3 k^2}{2\epsilon_0 c \beta^2 P^2}\Theta\left[|\vec{e}|^2|e_z|^2+\frac12\vec{e}^2(e_z^*)^2\right]\;,\\
\label{eq:gamD}
\gamma_D&=&\frac{k^2}{2\epsilon_0 c \beta^2P^2}\int_{-\infty}^{+\infty}\chi_3\left(|\vec{e}|^4+\frac12 |\vec{e}^2|^2\right)dx\;,\\
\label{eq:P}
P&=&\int_{-\infty}^{+\infty}|\vec{e}|^2 dx=\frac{2\beta^2-\epsilon_s}{2q_s^3}+\frac{2\beta^2-\epsilon_c}{2q_c^3}\;,
\end{eqnarray}
the surface-induced nonlinearity enhancement factor $g$  is \cite{SGM2011}:
\begin{eqnarray}
g&=&(1+\eta)^{-2}\;,\\
\eta&=&\frac{-i}{\beta P}\Delta[e_z^*e_x]=-\frac{1}{P}\left(\frac{1}{q_s}+\frac{1}{q_c}\right)\;,
\end{eqnarray}
and the effective linear absorption parameter is given by:
\begin{equation}
\Lambda=g^{1/2}\frac{\alpha_1^{(R)}k}{2\beta P}\Theta[|e_z|^2]\;.
\end{equation}
In the above derivations we used the auxiliraly relation $g^{1/2}Q=\beta^2 P$, which can be obtained by using $i\beta e_z=-\partial_x e_x$ and taking by parts integral in Eq.~(\ref{eq:P}) \cite{SGM2011}.

Expression for graphene nonlinear coefficient in Eq. (\ref{eq:gamG}) can be replaced by the integral similar to the one in Eq. (\ref{eq:gamD}), following introduction of an effective graphene nonlinear susceptibility:
\begin{equation}
\label{eq:chi_graph}
 \chi_3^{(gr)}=i\alpha_3 \delta(x)=\frac{i\sigma^{(3)}}{\epsilon_0 c}\delta(x)\;,
\end{equation}
 where $\delta(x)$ is the Dirac delta function. Note however the different structure of the term under the integral, which is due to the surface nature of nonlinear response in graphene, cf. Eqs.~(\ref{eq:surf_current}) and (\ref{eq:Kerr}). In the limit of high localization, $\beta\gg \epsilon_{s,c}$, for the guided mode one obtains simple relation $e_x=\pm i e_z$, and therefore:
\begin{equation}
|\vec{e}|^2|e_z|^2+\frac12\vec{e}^2(e_z^*)^2\approx \frac12 |\vec{e}|^4\;, \qquad \vec{e}^2\approx 0\;.
\end{equation}
Apparently, in this limit, the effective nonlinear response of graphene is twice weaker than that of a infinitesimally thin Kerr medium with the susceptibility $\chi_3^{(gr)}$.

\section{Quasi-TE surface plasmon}

For the case of quasi-TE mode we use the ansatz:
\begin{eqnarray}
E_x&=&\left[C_x(\psi,x) +O(s^{2})\right]e^{i\beta z}\;,\\
E_y&=&\left[A(\psi,x)+B(\psi,x)+O(s^{5/2})\right]e^{i\beta z}\;, \\
E_z&=&\left[C_z(\psi,x) +O(s^{2})\right]e^{i\beta z}\;,
\end{eqnarray}
where $\partial_z\psi\sim s$, $\partial_y\psi\sim s^{1/2}$, $A\sim s^{1/2}$, $C_{x,z}\sim s$, $B \sim s^{3/2}$. 
Following substitution into Maxwell's equations, in the order $O(s^{1/2})$ we obtain the following boundary value problem:
\begin{eqnarray}
\hat{L}_{TE} A&=&0\;,\\
\label{eq:BC_TE_order_s12}
\Delta[A]=0\;,&& \Delta[\partial_x A]=-\alpha_1^{(I)} \Theta[A]\;,\\
\label{eq:L_TE}
\hat{L}_{TE}&=&\beta^2-\epsilon-\partial^2_{xx}\;.
\end{eqnarray}

We choose the solution in the form $A=I^{1/2}\psi(z,y)e_y$, where $e_y$ is the surface plasmon mode:
\begin{eqnarray}
x<0&:& e_y=e^{q_sx}\;,\\
x>0&:& e_y=e^{-q_cx}\;,
\end{eqnarray}
$q_{s,c}$ are defined in Eq.~(\ref{eq:q_sc}), and the normalization factor $I$ ensures $|\psi|^2$ gives the power density carried in the $z$ direction:
\begin{eqnarray}
I&=&\frac{4 k}{\beta \epsilon_0 c P}\;,\\ 
P&=&\int_{-\infty}^{+\infty}  |e_y|^2 dx=\frac{1}{2q_s}+\frac{1}{2 q_c}\;.
\end{eqnarray}
Dispersion relation for the TE plasmon is given by:
\begin{equation}
\label{eq:TE_disp}
\sqrt{\beta^2-\epsilon_s}+\sqrt{\beta^2-\epsilon_c}=-\alpha_1^{(I)}\;.
\end{equation}

In the order $O(s)$ we obtain:
\begin{eqnarray}
\label{eq:TE_order_s}
\hat{L}_{TM} \vec{C}&=&-I^{1/2}\partial_y \psi [\partial_x e_y,i\beta e_y]^T\;,\\
\Delta[C_z]=0\;,&& \Delta[i\beta C_x - \partial_x C_z]=\alpha_1^{(I)} \Theta[C_z]\;,
\label{eq:BC_TE_order_s}
\end{eqnarray}
where $\vec{C}=[C_x,C_z]^T$, operator $\hat{L}_{TM}$ is defined in Eq.~(\ref{eq:L_TM}). Substituting in the above equations $\vec{C}=I^{1/2}\partial_y \psi\vec{e}$ and eliminating $e_x$, we obtain:
\begin{equation}
\hat{L}_{TE}e_z=i\beta \epsilon^{-1}\hat{L}_{TE}e_y=0\;.
\end{equation}
In other words, $e_z$ solves the same homogeneous equation as $e_y$. 
Comparing boundary conditions for $e_y$, Eq.~(\ref{eq:BC_TE_order_s12}), and $e_z$, 
Eq.~(\ref{eq:BC_TE_order_s}), we obtain $e_z=(i/\beta)e_y$ and $e_x\equiv 0$. It is easy to check that this choice also satisfies the condition $\vec\nabla\cdot \vec{E}=0$ in the order $O(s)$.

In the order $O(s^{3/2})$ the following boundary value problem is obtained:
\begin{eqnarray}
\label{eq:TE_order_s32}
&\hat{L}_{TE}\vec{B}=I^{1/2}\left[i2\beta\partial_z\psi e_y -\partial^2_{yy}\psi( i\beta e_z+\partial_x e_x)\right]\;,&\\
&\Delta[B]=0\;,&\\
\nonumber
&\Delta[\partial_x B - \partial_y C_x]=i\alpha_1^{(R)}I^{1/2}\psi\Theta[e_y]&\\
\label{eq:BC_TE_order_s32}
&\qquad\qquad-\alpha_1^{(I)}\Theta[B]+i\frac{3}{4}\alpha_3 I^{3/2}|\psi|^2\psi\Theta[|e_y|^2 e_y]\;,&
\end{eqnarray}
Projecting Eq.~(\ref{eq:TE_order_s32}) onto the mode $e_y$ and following essentially the same steps as described in the previous section, in the l.h.s. we obtain:
\begin{eqnarray}
\int_{-\infty}^{+\infty}e_y^*\hat{L}_{TE} B dx=-\Theta[e_y^* \partial_x B - B\partial_x e_y^*]\;.
\end{eqnarray}
Performing projection in the r.h.s. of Eq.~(\ref{eq:TE_order_s32}), and using boundary conditions in Eqs.~(\ref{eq:BC_TE_order_s12}) and (\ref{eq:BC_TE_order_s32}), we obtain the amplitude equation (\ref{eq:NLS}) with the following coefficients:
\begin{eqnarray}
\label{eq:Upsilon_TE}
\Upsilon_{TE}&=&\gamma_{G,TE}+\gamma_{D,TE}\;,\\
\label{eq:gamG_TE}
\gamma_{G,TE}&=&\frac{i3\alpha_3 k^2}{2\epsilon_0 c \beta^2 P^2}\Theta\left[|e_y|^4\right]\;,\\
\label{eq:gamD_TE}
\gamma_{D,TE}&=&\frac{3k^2}{2\epsilon_0 c \beta^2P^2}\int_{-\infty}^{+\infty}\chi_3|e_y|^4 dx\;,\\
\Lambda_{TE}&=&\frac{\alpha_1^{(R)}k}{2\beta P}\Theta[|e_y|^2]\;.
\end{eqnarray}

For quasi-TE mode the enhancement factor $g$ is absent. Also, due to the linear mode being scalar, in this case the nonlinear response of graphene is completely analogoues to that of a infinitesimally thin Kerr medium with the susceptibility $\chi_3^{(gr)}$ in Eq.~(\ref{eq:chi_graph}).

\section{Analysis and discussion}

\begin{figure}
\includegraphics[width=0.4\textwidth]{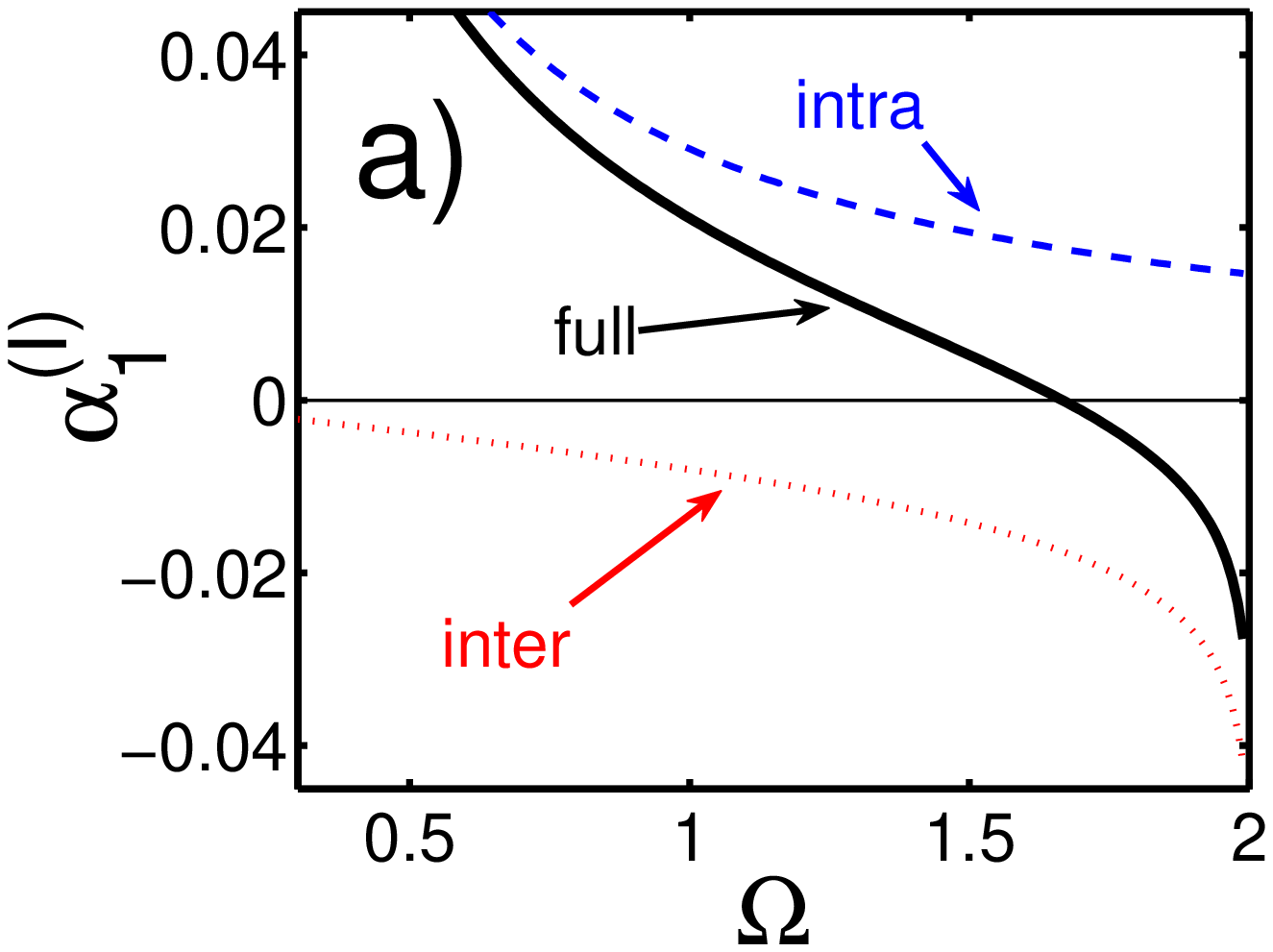}
\includegraphics[width=0.4\textwidth]{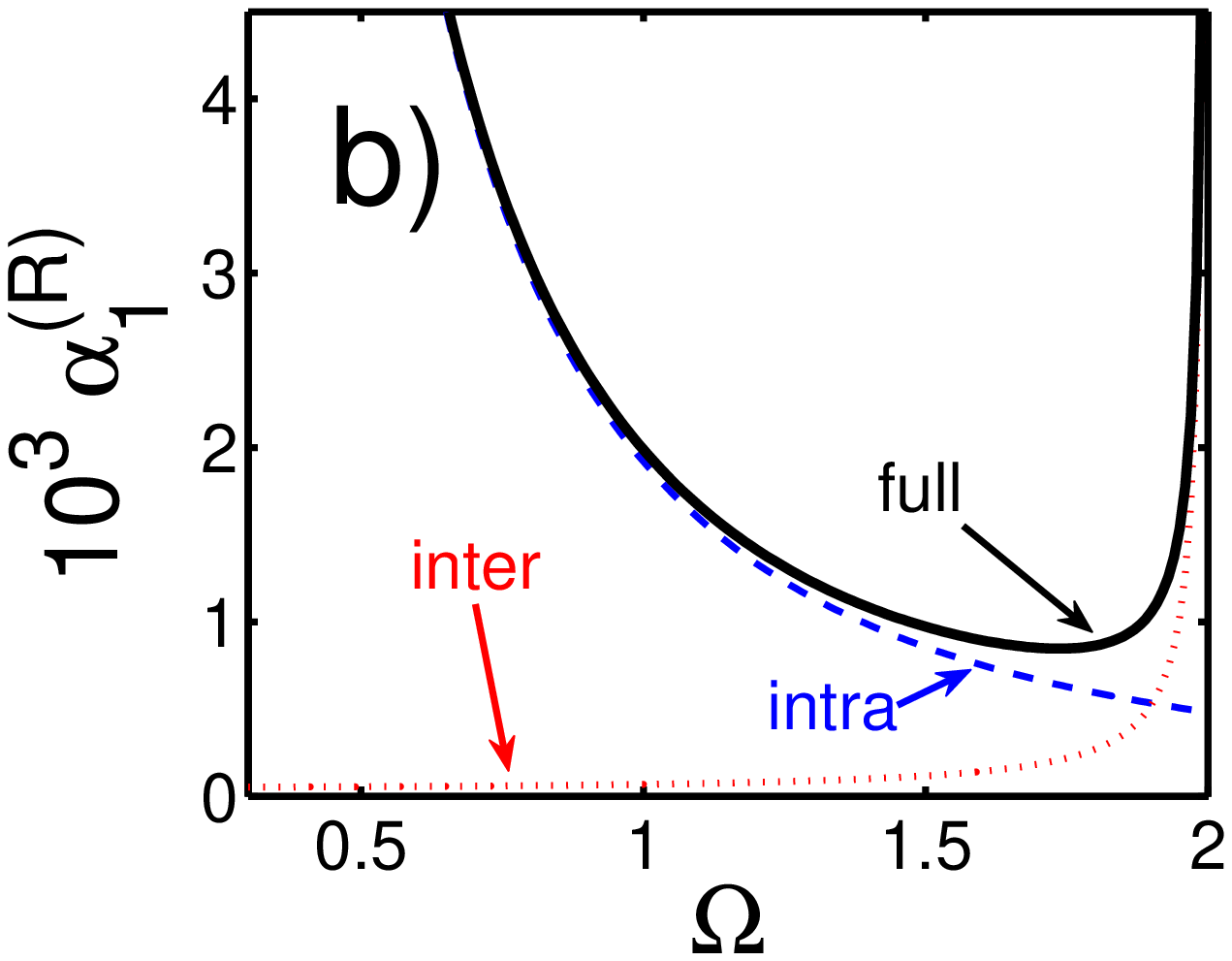}
\caption{(Color online) Dimensionless conductivity of graphene $\alpha_1$: imaginary (a) and real (b) parts. Dashed, dotted and solid curves correspond to intra-, inter- and full conductivity, respectively. The chemical potential is set to $\mu=0.1$eV, and relaxation times are $\tau_{intra}=100$fs, $\tau_{inter}=1$ps.}
\label{fig:sigmas}
\end{figure}

Conductivity of graphene consists of intra- and inter-band contributions,  $\sigma_1=\sigma_{intra}+\sigma_{inter}$. For the case of a highly doped graphene, 
$|\mu|\gg kT$, $\mu$  is the chemical potential, intra- and inter-band terms are given by the semi-classical formalism \cite{Mikhailov2007a}:
\begin{eqnarray}
\sigma_{intra}(\Omega)&=&\frac{ie^2}{\pi \hbar}\cdot \frac{1 }{\Omega +i\nu_{intra}}\;,\\
\sigma_{inter}(\Omega)&=&\frac{ie^2}{4\pi\hbar}\ln\frac{2-|\Omega|-i\nu_{inter}}{2+|\Omega|+i\nu_{inter}}\;,
\end{eqnarray}
where $\Omega=\hbar\omega/\mu$, excitation below interband absorption threshould is assumed: $\Omega<2$, coefficients $\nu=\hbar/(|\mu|\tau)$ take into account losses due to electron scatterings at finite temperatures, below we take $\tau_{intra}=100$fs and $\tau_{inter}=1$ps \cite{Jablan2009, Gu2012}. 
For the doping level of $\mu=0.1$eV we obtain $\nu_{intra}\approx 0.066$, $\nu_{inter}\approx 0.007$, and the interband absorption threshould is at 
$\omega_{th}=2\mu/\hbar\approx 3\cdot 10^{14}$ rad/s ($\lambda_{th}\approx 6.3\mu$m). The corresponding dimensionless conductivity $\alpha_1$ is plotted in Fig.~\ref{fig:sigmas}. Imaginary part of $\alpha_1$ changes its sign at $\Omega_0\approx 1.67$, linear TM plasmons exist for $\Omega<\Omega_0$ (i.e. when $\alpha_1^{(I)}>0$), while TE -- for $\Omega>\Omega_0$ ($\alpha_1^{(I)}<0$) \cite{Mikhailov2007a}. For the chosen doping level, $\Omega_0$ corresponds to $\lambda_0\approx 7.5\mu$m.

The nonlinear conductivity coefficient $\sigma_3$ for graphene is given by \cite{Mikhailov2008}:
\begin{equation}
\sigma_3(\Omega)=-i\frac{3}{32}\frac{e^2}{\pi \hbar}\frac{(eV_F)^2\hbar^2}{\mu^4 \Omega^3}(1+i\alpha_{T})\;.
\label{eq:sigma_3}
\end{equation}
Here we introduced coefficient $\alpha_{T}$ to account for two photon absorption in graphene, recent experiments suggest $\alpha_{T}\approx 0.1$ \cite{Gu2012}. Negative imaginary part of $\sigma_3$ suggests that the nonlinear response is of self-focusing type, cf. Eqs. (\ref{eq:NLS}), (\ref{eq:gamG}) and (\ref{eq:gamG_TE}).

Below we consider surface plasmons in configurations with air ($\epsilon=1$) and silicon ($\epsilon=12$) as dielectrics. For silicon we take $\chi_3=(4/3)c\epsilon_0 \epsilon_s n_2$, $n_2=4\cdot 10^{-18} m^2/W$. Two photon absorption in silicon is negligible  for $\lambda>2\mu$m \cite{Bristow2007}. For simplicity we neglect dispersion of linear and nonlinear dielectric constants.

\subsection{Quasi-TM plasmon}

\begin{figure}
\includegraphics[width=0.4\textwidth]{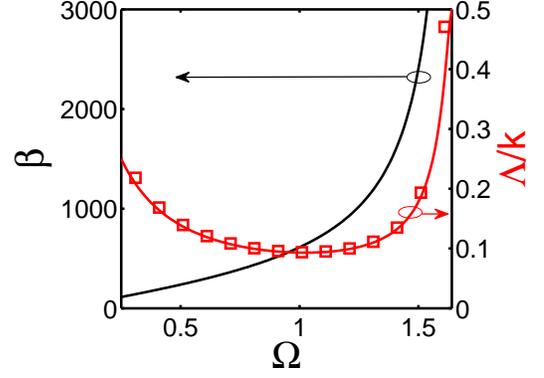}
\caption{(Color online) Dispersion of TM plasmon in the silicon-graphene-air geometry: propagation constant $\beta$ and loss parameter $\Lambda/k$ as functions of $\Omega$. Imaginary part of $\beta$ as computed from the full dispersion relation with complex-valued $\alpha_1$ is shown with squares.}
\label{fig:disp_TM}
\end{figure}

First, we consider TM plasmon in the configuration with silicon substrate and air cladding. The corresponding dispersion is plotted in Fig.~\ref{fig:disp_TM}. To validate our theory, the propagation loss parameter $\Lambda/k$ is compared against imaginary part of the propagation constant (see open squares in Fig.~\ref{fig:disp_TM}). The latter is computed from the full dispersion relation that takes into account complex-valued $\alpha_1$ and obtained by  replacing $\alpha_1^{(I)}$ with $-i\alpha_1$ in Eq.~(\ref{eq:disp_TM}). The results are found to be in perfect agreement.

TM plasmon is characterized by a considerable light confinement in a wide range of frequencies: even at the frequency as low as $\Omega=0.25$ 
($\omega=7.5\cdot 10^{13}$rad/s, $\lambda\approx 25\mu$m) the propagation constant is $\beta\approx 240$, and it constantly grows as $\Omega$ increases towards the threshould value $\Omega_0$. Propagation losses are relatively low: $\Lambda/(k\beta)<10^{-3}$ when $0.5<\Omega<1.6$,
which is due to the low absorption rate in graphene below the interband absorption threshould, cf. Fig.~\ref{fig:sigmas}(b).

\begin{figure}
\includegraphics[width=0.4\textwidth]{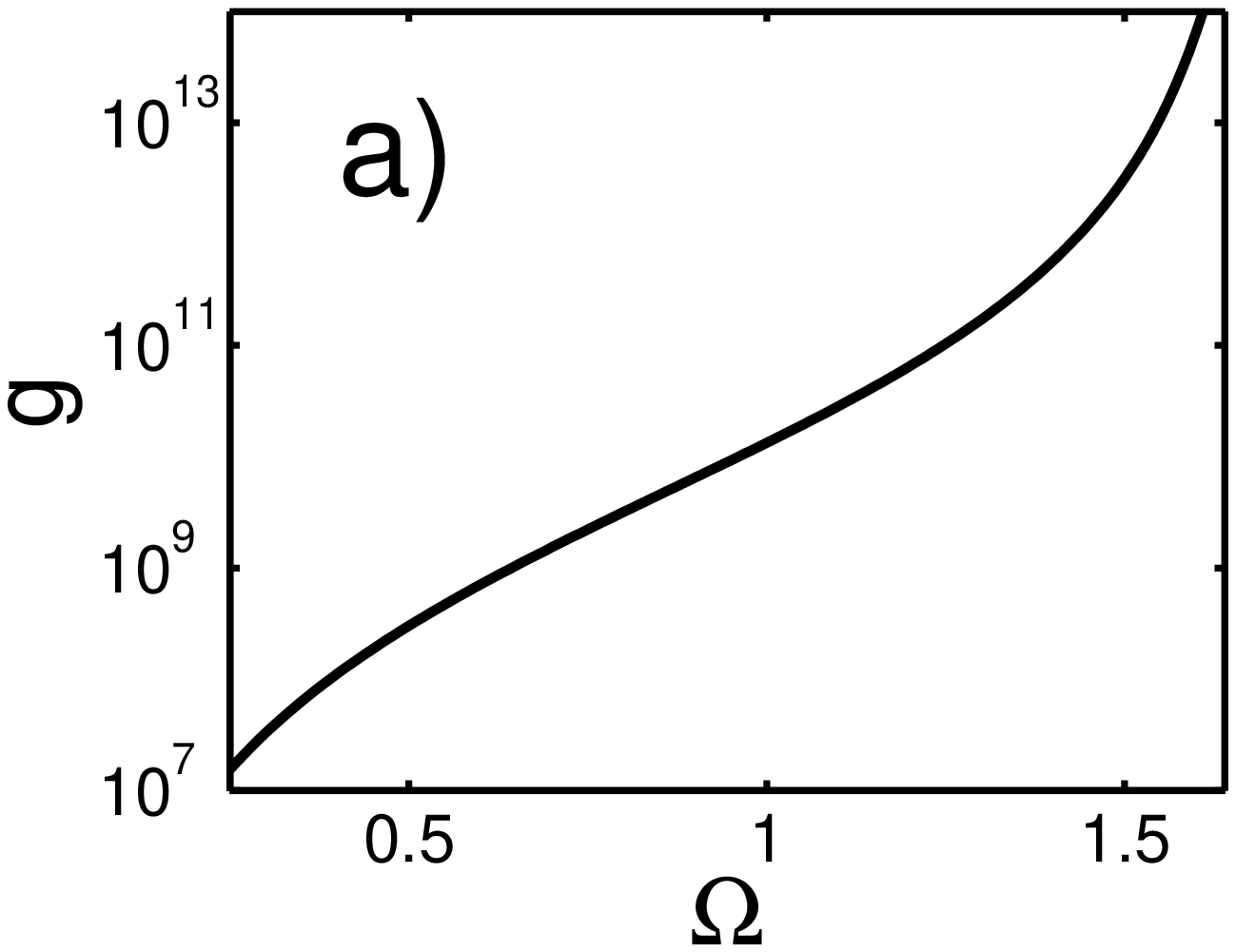}
\includegraphics[width=0.4\textwidth]{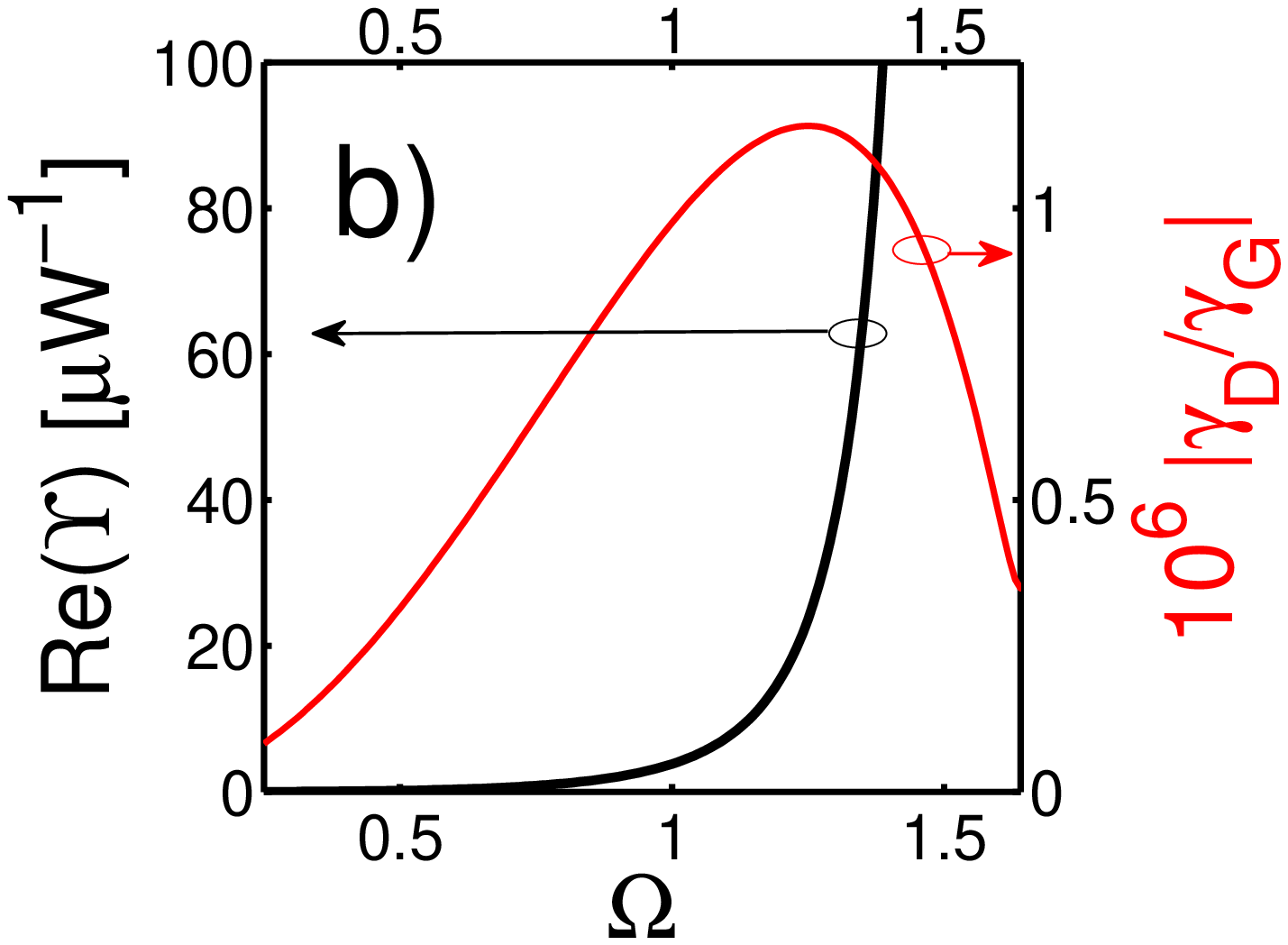}
\caption{(Color online) Effective nonlinearity of quasi-TM plasmon: (a) surface enhancement factor $g$; (b) nonlinear coefficient $\Upsilon$ and the relative dielectric nonlinearity.}
\label{fig:NL_TM}
\end{figure}

Due to the high localization of TM plasmon, the surface-induced enhancement factor $g$ is large and is growing nearly exponentially with increasing $\Omega$, see Fig.~\ref{fig:NL_TM}(a). This growth over-ballances the decay of graphene nonlinear response $\sigma_3\sim \Omega^{-3}$, cf. Eq.~(\ref{eq:sigma_3}), and causes 
the considerable increase of the nonlinear coeffficient $\Upsilon$ with frequency, see Fig.~\ref{fig:NL_TM}(b). 

\begin{figure}
\includegraphics[width=0.4\textwidth]{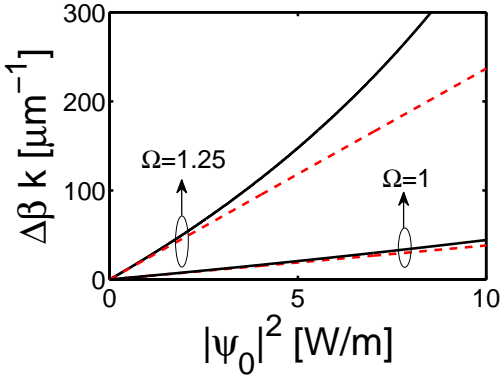}
\caption{(Color online) Nonlinear index shift $\Delta\beta=\beta_{NL}-\beta$ vs power density. Solid curves correspond to the numerical solution of the dispersion relation in Eq.~(\ref{eq:beta_NL_anal}), dashed lines - to the result given by the amplitude Eq. (\ref{eq:NLS}): $\Delta\beta k=\Upsilon|\Psi_0|^2$.}
\label{fig:beta_shifts_TM}
\end{figure}

Remarkably, the relative contribution of silicon substrate to the overal nonlinearity remains negligibly small within the entire frequency window of existence of TM plasmon, as illustrated in Fig.~\ref{fig:NL_TM}(b). Also, due to the large $\beta$, the diffraction term in Eq.~(\ref{eq:NLS}) can be neglected for typical beam widths $L_y>1\mu m$. Indeed, for $\Omega=1$ ($\lambda\approx 12.4\mu m$) and the beam width of $L_y=10\mu m$ the diffraction length is $L_D=L_y^2\beta k\approx 50$mm, which is more than six orders of magnitude larger that the apparent plasmon wavelength $\lambda_p=2\pi/(\beta k)\approx 10$nm. Neglecting nonlinear response of dielectrics and beam diffraction, as well as disregarding linear and nonlinear absorption ($\sigma_1^{(R)}=\sigma_3^{(R)}=0$), one can find stationary solutions of Maxwell equations $\vec{E}(x,y,z)=I\psi_0 \vec{e}(x,y;\beta_{NL})e^{i\beta_{NL} z}$ with the nonlinear boundary condition in Eqs.~(\ref{eq:BC_Ey_and_Ez}-\ref{eq:BC_Hz}) analytically. The corresponding dispersion relation reads:
\begin{eqnarray}
\nonumber
&\frac{\epsilon_s}{\sqrt{\beta_{NL}^2-\epsilon_s}}+\frac{\epsilon_c}{\sqrt{\beta_{NL}^2-\epsilon_c}}=\alpha_1^{(I)}&\\
\label{eq:beta_NL_anal}
&\qquad\qquad+\frac{\alpha_3^{(I)}}{2}I\left[|\vec{e}|^2|e_z|^2+\frac12\vec{e}^2(e_z^*)^2\right]|\psi_0|^2&\;.
\end{eqnarray}
At the same time, substituting $\psi(y,z)\equiv \psi_0e^{i\beta_{NL}z}$ into the amplitude equation (\ref{eq:NLS}) and assuming $\Lambda=Im(\Upsilon)=0$, we obtain $\beta_{NL}=\beta+(\Upsilon/k)|\psi_0|^2$. Solving the dispersion relation in Eq.~(\ref{eq:beta_NL_anal}) numerically, we found both results to be in perfect agreement at low power densities $|\psi_0|^2$, see Fig.~\ref{fig:beta_shifts_TM}.

\subsection{Quasi-TE plasmon}

\begin{figure}
\includegraphics[width=0.4\textwidth]{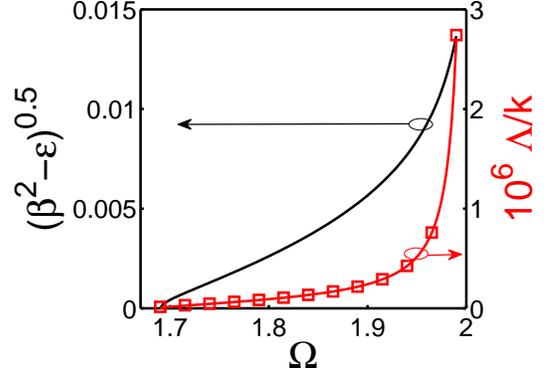}
\caption{(Color online) Dispersion of TE plasmon in the silicon-graphene-silicon geometry: 
Plasmon localization factor $q=\sqrt{\beta^2-\epsilon}$, propagation loss $\Lambda/k$ and imaginary part of the propagation constant (squares).}
\label{fig:disp_TE}
\end{figure}

As follows from the dispersion relation in Eq.~(\ref{eq:TE_disp}), in order to excite TE plasmon in an assymetric geometry with different substrate and cladding dielectrics one has to make conductivity of graphene strong enough: $|\alpha^{(I)}|>\sqrt{|\epsilon_s-\epsilon_c|}$. We would like to note however, in contrast to the case of a conventional dielectric slab waveguide, here only the {\em difference} between the cladding and substrate dielectric constants matters. 
Apparently, for the chosen doping level of graphene, TE plasmon does not exist in the silicon-graphene-air configuration, cf. Fig.~\ref{fig:sigmas}(a). Instead, we consider the fully symmetric configuration with silicon in the cladding and substrate: $\epsilon_s=\epsilon_c=12$. The corresponding dispersion is plotted in Fig.~\ref{fig:disp_TE}. Due to the low values of $|\alpha^{(I)}|$, TE plasmon is only weakly localized. However, one benefits from much smaller propagation losses per plasmon period $T=2\pi/(\beta k)$, compared to TM plasmons.

\begin{figure}
\includegraphics[width=0.4\textwidth]{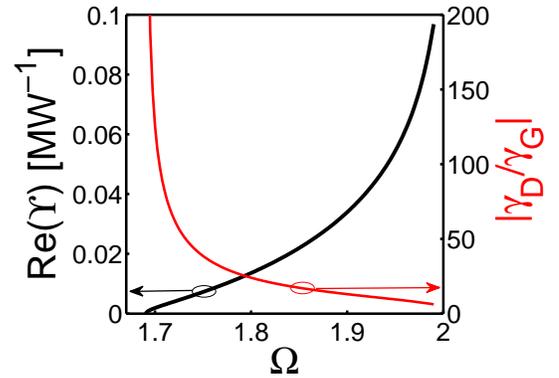}
\caption{(Color online) The same as Fig.~\ref{fig:NL_TM}(b) but for quasi-TE plasmon in the silicon-graphene-silicon geometry.}
\label{fig:NL_TE}
\end{figure}

As the result of weak localization, typical values of the effective nonlinear coefficient $\Upsilon$ for TE plasmons are nearly twelve orders of magnitude below those for TM plasmons, see Fig.~\ref{fig:NL_TE}. Remarkably, for TE plasmon nonlinear contribution from the dielectrics is dominant over graphene, they become comparable only when the frequency $\Omega$ approaches the interband absorption threshould $\Omega_{th}=2$.

With the account of small propagation constants $\beta$, diffraction term and associated effects due to its interplay with the focusing nonlinearity become important for TE plasmons with typical widths of the order of several micrometres. Taking $\Omega=1.9$ (corresponding to $\lambda\approx 6.5\mu$m),  for $L_y=5\mu$m the diffraction length becomes $L_D=L_y^2\beta k\approx 30\mu$m, and the plasmon period is $2\pi/(\beta k)\approx 0.5\mu$m. However, due to the low nonlinearity, one requires considerably high powers to observe basic effects such as self-focusing. For instance, to form a spatial soliton of width $L_y$ the peak power density should be $|\psi_0|^2=(\Upsilon L_D)^{-1}\approx 3\cdot 10^{10} W/m$, and therefore the total power must be of the order of several mega-Watts.

\section{Summary}

Using asymptotic expansion of Maxwell equations and boundary conditions, we have derived amplitude equation for nonlinear TM and TE surface plasmon waves in the dielectric-graphene-dielectric planar configuration. Induced surface current in graphene shows strongly nonlinear response to the applied electromagnetic field. We have shown that this leads to the effective focussing Kerr type nonlinearity. For TE plasmon this nonlinearity is fully analogous to that of an infitesimally thin dielectric layer. However, for TM plasmon the structure of the corresponding nonlinear coefficient is different and reflects the unique surface-only response of graphene.

For typical doping levels of graphene of the order of $0.1$eV, TM plasmons are strongly localized. 
This causes the significant enhancement of nonlinearity, we predict that considerable nonlinear phase shifts can be observeed for power densities as low as few micro-Watts per metre. Remarkably, graphene contribution to the overall nonlinearity is shown to be strongly dominant over that of dielectrics in this case. In contrast, TE plasmons are only weakly localized, and the major part of the overall nonlinearity is due to dielectric substrate and cladding. Typical values of the nonlinear coefficient for TE plasmons are found to be about 12 orders of magnitude below those for TM plasmons.

\begin{acknowledgements}
We acknowledge usefull discussions with Dmitry Skryabin.
\end{acknowledgements}

\bibliographystyle{apsrevmy}

%\bibliography{Graphene,mypapers,somebooks,soi}

\end{document}